\newcommand{\vtr}[1]{\boldsymbol{#1}}
\newcommand{\fig}[1]{Fig.~\ref{#1}}
\newcommand{\Fig}[1]{Fig.~\ref{#1}}

\def\mupara{\boldsymbol{\mu }_{\text{para}}}
\def\Hext{\boldsymbol{H}}
\def\Mexp{\boldsymbol{M}_\text{exp}}

\def\rhot{\rho^{\text{SMR Trans}}_{\text{sim}}}
\def\rhote{\rho^{\text{SMR Trans}}_{\text{exp}}}

\documentclass[reprint,
superscriptaddress,
amsmath,amssymb,
 aps,
prb
]{revtex4-2}

\usepackage{graphicx}\usepackage{dcolumn}\usepackage{bm}\usepackage{chemformula}\usepackage{multirow}
\usepackage{array}
\usepackage{verbatim}
\usepackage[normalem]{ulem} \usepackage[colorlinks=true,citecolor=blue,linkcolor=magenta, breaklinks=true]{hyperref}
\usepackage{upgreek}
\newcolumntype{P}[1]{>{\centering\arraybackslash}p{#1}}
\usepackage[colorlinks=true,citecolor=blue,linkcolor=magenta]{hyperref}

\begin{document}

\preprint{APS/PRB}

\title{Probing magnetic anisotropy and spin-reorientation transition in 3D antiferromagnet, \ch{Ho_{0.5}Dy_{0.5}FeO_{3}}$\vert$\ch{Pt} using spin Hall magnetoresistance}

\author{Aditya A. Wagh}
\email[]{adityawagh@iisc.ac.in}
\affiliation{Department of Physics, Indian Institute of Science, Bangalore, INDIA }

\author{Priyanka Garg}
\email[]{priyankagarg@iisc.ac.in}
\affiliation{Department of Physics, Indian Institute of Science, Bangalore, INDIA }

\author{Arijit Haldar}
\affiliation{Department of Physics, Indian Institute of Science, Bangalore, INDIA }\affiliation{Department of Physics, University of Toronto, 60 St. George Street, Toronto, ON, M5S 1A7 Canada}

\author{Kingshuk Mallick}

\author{Tirthankar Chakraborty}

\author{Suja Elizabeth}
\affiliation{Department of Physics, Indian Institute of Science, Bangalore, INDIA }

\author{P. S. Anil Kumar}
\affiliation{Department of Physics, Indian Institute of Science, Bangalore, INDIA }

\begin{abstract}

Orthoferrites ($RE$\ch{FeO_{3}}) containing rare-earth ($RE$) elements are 3D antiferromagnets  (AFM) that exhibit characteristic weak ferromagnetism originating due to slight canting of the spin moments and display a rich variety of spin reorientation transitions in the magnetic field ($H$)-temperature ($T$) parameter space. We present spin Hall magnetoresistance (SMR) studies on a $b$-plate ($\mathit{ac}$-plane) of crystalline \ch{Ho_{0.5}Dy_{0.5}FeO_{3}}$\vert$\ch{Pt} hybrid at various $T$ in the range, 11 to 300 K. In the room temperature $\Gamma_{4}(G_{x}, A_{y}, F_{z})$ phase, the switching between two degenerate domains, $\Gamma_{4}(+G_{x}, +F_{z})$ and $\Gamma_{4}(-G_{x}, -F_{z})$ occurs at fields above a critical value, $H_{\text{c}} \approx 713$ Oe. Under $H > H_{\text{c}}$, the angular dependence of SMR ($\alpha$-scan) in the $\Gamma_{4}(G_{x}, A_{y}, F_{z})$ phase yielded a highly skewed curve with a sharp change (sign-reversal) along with a rotational hysteresis around $\mathit{a}$-axis. This hysteresis decreases with an increase in $H$. Notably, at $H < H_{\text{c}} $, the $\alpha$-scan measurements on the single domain, $\Gamma_{4}(\pm G_{x}, \pm F_{z})$ exhibited an anomalous sinusoidal signal of periodicity 360 deg. Low-$T$ SMR curves ($H$ = 2.4 kOe), showed a systematic narrowing of the hysteresis (down to 150 K) and a gradual reduction in the skewness (150 to 52 K), suggesting weakening of the anisotropy possibly due to the $T$-evolution of \ch{Fe}-$RE$ exchange coupling. Below 25 K, the SMR modulation showed an abrupt change around the $\mathit{c}$-axis, marking the presence of $\Gamma_{2}(F_{x},C_{y},G_{z})$ phase. We have employed a simple Hamiltonian and computed SMR to examine the observed skewed SMR modulation. In summary, SMR is found to be an effective tool to probe magnetic anisotropy as well as a spin reorientation in \ch{Ho_{0.5}Dy_{0.5}FeO_{3}}. Our spin-transport study highlights the potential of \ch{Ho_{0.5}Dy_{0.5}FeO_{3}} for future AFM spintronic devices.

\end{abstract}

\maketitle

\section{\label{sec:level1}INTRODUCTION}

Antiferromagnetic (AFM) insulators are envisaged as the next-generation memory devices to exploit switching of the N\'eel vector. They offer advantages over ferromagnets (FM), such as zero macroscopic magnetization, enhanced robustness against magnetic field ($\Hext$) fluctuations and the faster switching dynamics \cite{jungwirth2016antiferromagnetic, wadley2016electrical, cheng2020electrical}. 
In recent years, the potential of AFM passive layers has been highlighted in various spintronic devices. For instance, in the experimental demonstration of a magnon valve, a passive insertion-layer of AFM \ch{NiO} was used to weaken magnetostatic coupling between the two adjacent \ch{Y_{3}Fe_{5}O_{12}} (YIG) layers while allowing the magnon transport \cite{guo2018magnon}. On the other hand, the passive AFM layers are found to play a key role in pinning the adjacent FM layers in some spintronic devices \cite{hou2017tunable}. Moreover, exotic spin configurations in AFMs have been seen to absorb spin currents at AFM$\vert$heavy metal (HM) interfaces resulting in interesting modulations in spin Hall magnetoresistance (SMR) \cite{wang2017antiferromagnetic, aqeel2016electrical}. AFMs are observed to exhibit characteristic negative SMR (out-of-phase modulation) compared to the conventional positive SMR in collinear FMs \cite{fischer2018spin, geprags2020spin, nakayama2013spin, chen2013theory, chen2016theory}. Recent studies on FM$\vert$AFM$\vert$HM hybrid structures have highlighted the high tunability of (non-local) SMR signal in such devices \cite{hoogeboom2021role, lin2017electrical, hou2017tunable, shang2016effect}. 3D AFM materials exhibit very rich magnetic phase diagrams and hence, are a good prospect for studying the spin transport phenomena. Further, SMR at AFM$\vert$HM interface can be a useful probe to investigate various magnetic phases and the phase transitions in 3D AFMs.

SMR studies on epitaxial thin film \ch{SmFeO_{3}}$\vert$\ch{Ta} bilayer showed that the sign of the SMR was positive and an ordering of the \ch{Sm}-sublattice enhanced its magnitude at low temperatures \cite{hajiri2019spin}. In another investigation on epitaxial film, \ch{TmFeO_{3}}$\vert$\ch{Pt}, the SMR was particularly measured in the transversal geometry (spin Hall-induced anomalous Hall effect (SHAHE)) to probe the spin reorientation transition (SRT) from $\Gamma_{4}$ to $\Gamma_{2}$ phase \cite{becker2021electrical}. In a recent report on the single crystal hybrid \ch{DyFeO_{3}}$\vert$\ch{Pt}, the angular dependence of SMR in $\Gamma_{4}$ phase was examined, while rotating the $\Hext$ in $\mathit{ab}$-plane. The observed sharp anomalies in SMR were analysed using a simple phenomenological model and attributed to the sudden change in the N\'eel vector \cite{hoogeboom2021magnetic}. 

In \ch{DyFeO_{3}} \ch{Fe}-sublattice undergoes an SRT, $\Gamma_{4} \xrightarrow{50 K} \Gamma_{1}$ whereas, \ch{HoFeO_{3}} exhibits $\Gamma_{4} \xrightarrow{58 K} \Gamma_{412} \xrightarrow{50 K} \Gamma_{2}$ with an intermediate transition region, $\Gamma_{412}$ \cite{mohammed2019magnetic,li2019spin,shao2011single}. Alternatively, some neutron studies on single crystal \ch{HoFeO_{3}} claimed that there are two SRTs; $\Gamma_{4} \xrightarrow{55 K} \Gamma_{1} \xrightarrow{36 K} \Gamma_{2}$ \cite{ovsyanikov2020neutron}. Chakraborty \textit{et al} \cite{chakraborty2018two} reported the presence of two-fold SRT, in their magnetic studies on single crystals of \ch{Ho_{0.5}Dy_{0.5}FeO_{3}} i.e. $\Gamma_{4} \xrightarrow{50 K} \Gamma_{1} \xrightarrow{26 K} \Gamma_{2}$. In the present study, we investigate spin transport at single crystal \ch{Ho_{0.5}Dy_{0.5}FeO_{3}}$\vert$\ch{Pt} interface. We examined the two-fold SRT in \ch{Ho_{0.5}Dy_{0.5}FeO_{3}} using SMR as our measurement probe. 

 In $\Gamma_{4}$ phase at room temperature (RT) both the basis vectors, $\vtr{G}_{\ch{Fe}}$ (the N\'eel vector) and $\vtr{F}_{\ch{Fe}}$ (weak ferromagnetism) lie in the $\mathit{ac}$-plane. Therefore, we chose a $\mathit{b}$-plate of \ch{Ho_{0.5}Dy_{0.5}FeO_{3}} single crystal and studied angular dependence of SMR while rotating the $\Hext$ in $\mathit{ac}$-plane. Our choice of the plane for $\Hext$-rotation, which is different from that of $\mathit{ab}$-plane in a recent report on \ch{DyFeO_{3}}, yielded some interesting results \cite{hoogeboom2021magnetic}. Angle-dependent SMR modulation revealed a sharp anomaly near the $\mathit{a}$-axis which was notably accompanied by a rotational hysteresis. In order to investigate the SMR, we employed a simple model considering competing interactions acted upon \ch{Fe} spins that include the nearest-neighbour and the next-nearest-neighbour exchange, Dzyaloshinskii–Moriya (DM) interaction, anisotropy and Zeeman energy. Notably, additional paramagnetic contribution from rare-earth sublattice was essential to account for the observed magnetization and SMR. 

In the $\Gamma_{4}$ phase, a degeneracy of \ch{Fe}-spin configuration (or domains), characterized by $\Gamma_{4}(+G_{x}, +F_{z})$ and $\Gamma_{4}(-G_{x}, -F_{z})$, allows achieving a single domain of the choice by applying $H$ above a critical value ($H_{c}$). Our low field ($H < H_{c}$) SMR measurements on such single domain yielded a signal with anomalous periodicity of $360\text{ deg}$, compared to the typical periodicity of $180\text{ deg}$ in the conventional SMR signals. Next, we trace the SRTs $\Gamma_{4} \xrightarrow{49 K} \Gamma_{1} \xrightarrow{26 K} \Gamma_{2}$, by carrying out SMR measurements at low temperatures. In $\Gamma_{4}$ phase at lower temperatures, an overall reduction in the skewness of the SMR-modulation was observed. This points towards the weakening of the anisotropy and can have possible origin in the presence of \ch{Fe}-$RE$ exchange coupling and its temperature evolution. In this report we have discussed in details, the effectiveness of SMR as a tool to examine magnetic anisotropy as well as spin reorientation in 3D AFM, \ch{Ho_{0.5}Dy_{0.5}FeO_{3}}.

\section{Magnetic Phases and Spin Re-orientation in $\text{Ho}_{0.5}\text{Dy}_{0.5}\text{FeO}_3$}
\label{Magnetism}
\begin{figure*}
\includegraphics[width=1\textwidth,keepaspectratio]{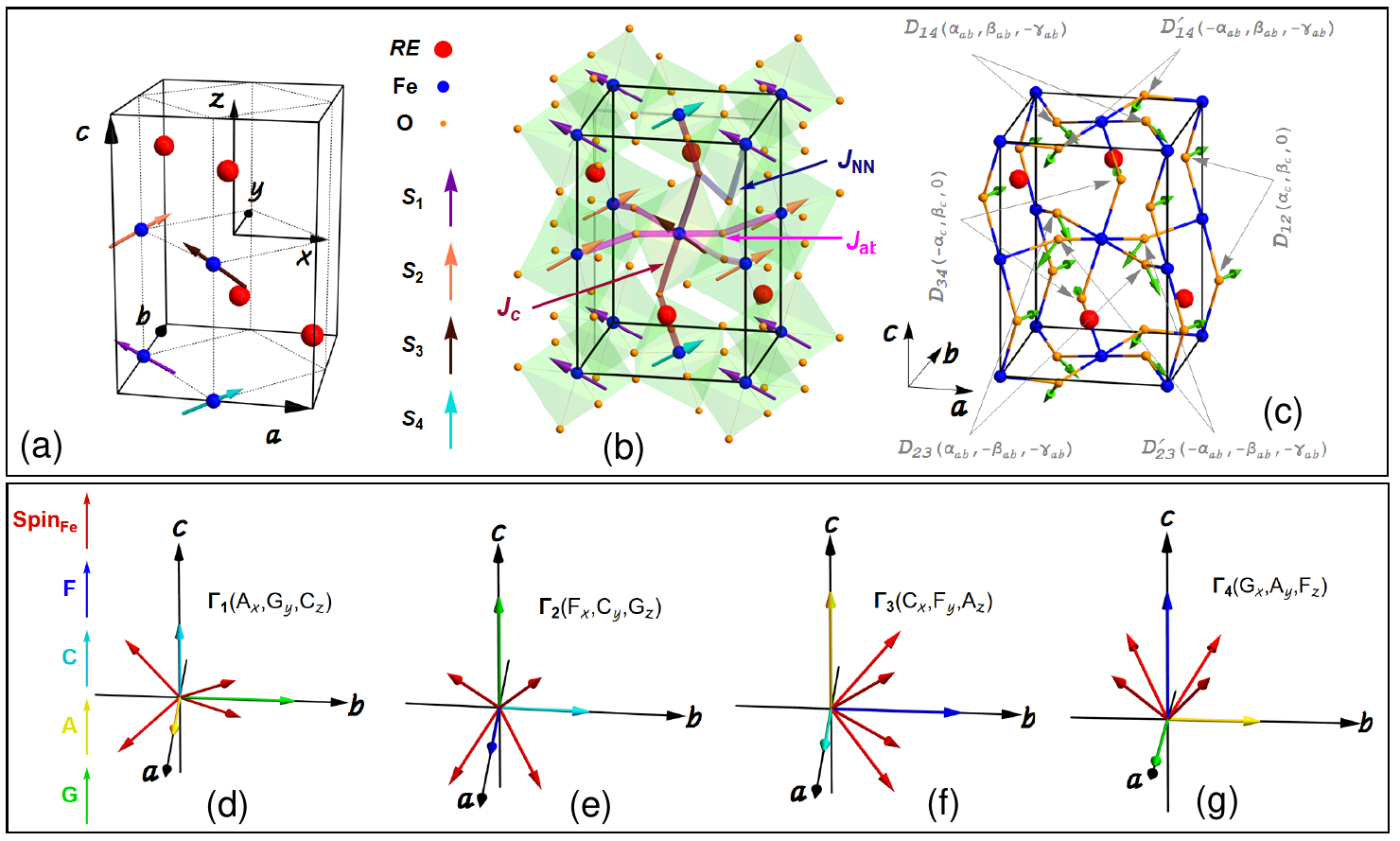}
\caption{\label{fig:1sch} Magnetic unit cell of perovskite orthoferrite, $RE$\ch{FeO_{3}}. (a) Orthorhombic unit cell with lattice vectors, $\vtr{a}$, $\vtr{b}$ and $\vtr{c}$. Four \ch{Fe^{3+}} spin moments are located at special positions. (b) Two types of exchange interactions between \ch{Fe^{3+}} spin moments; 1) nearest neighbour interaction ($J_{ab}$ and $J_{c}$), and 2) next-nearest neighbour interaction ($J_{NN}$). (c) Various Dzyaloshinskii–Moriya vectors are represented with green arrows. (d) to (g) Different representations of Pbnm space group such as $\Gamma_{1}$, $\Gamma_{2}$, $\Gamma_{3}$ and $\Gamma_{4}$ along with \ch{Fe^{3+}} spin moments and their respective basis vectors ($\vtr{F}$, $\vtr{C}$, $\vtr{A}$ and $\vtr{G}$).}
\end{figure*}

Orthoferrite \ch{Ho_{0.5}Dy_{0.5}FeO_{3}} exhibits distorted perovskite crystal structure with orthorhombic space group \textit{P}bnm ($\mathit{a}$ = 5.292 Å, $\mathit{b}$ = 5.591 Å and $\mathit{c}$ = 7.614 Å) \cite{chakraborty2018two}. Figure \ref{fig:1sch}(a) represents such orthorhombic unit cell with lattice vectors,  $\vtr{a}$, $\vtr{b}$ and $\vtr{c}$. A pseudo-cubic perovskite cell is shown with dashed lines for reference. The orthorhombic unit cell consists of four \ch{Ho^{3+}}/\ch{Dy^{3+}}, four \ch{Fe^{3+}} and twelve \ch{O^{2-}} ions. Magnetic ions (\ch{Ho^{3+}}/\ch{Dy^{3+}} and \ch{Fe^{3+}}) sit at the inequivalent positions as shown in Fig.\ref{fig:1sch}(a) \cite{yamaguchi1974theory}. Among \ch{Fe^{3+}} and \ch{Ho^{3+}}/\ch{Dy^{3+}} magnetic sub-lattices, the former sublattice comprises of $\vtr{S}_{1}, \vtr{S}_{2}, \vtr{S}_{3},  \vtr{S}_{4}$ spin moments and is represented by four arrows of distinct colors. Notably, an alternative system ($x y z$) is also shown in which $\vtr{x}$, $\vtr{y}$ and $\vtr{z}$ are parallel to $\vtr{a}$, $\vtr{b}$ and $\vtr{c}$, respectively. 

In a typical orthoferrite, magnetism is complex due to various competing superexchange interactions such as intra sublattice interactions (denoted by constants, $J^{\ch{Fe-Fe}}$ and $J^{RE-RE}$) and inter sublattice interaction (denoted by constant, $J^{\ch{Fe}-RE}$). Generally, \ch{Fe}-sublattice orders at high temperature ($T_{N}^{\ch{Fe}} >$ RT) and hence, $J^{\ch{Fe-Fe}}$ interactions are prominent. Similarly, $RE$-sublattice orders at very low temperature ($T_{N}^{RE}<$  10 K) and consequently, $J^{RE-RE}$ interactions play an important role at these low temperatures. However,  $RE$-sublattice is known to contribute to magnetism above $T_{N}^{RE}$ via weak $J^{\ch{Fe}-RE}$ interactions. In \ch{Ho_{0.5}Dy_{0.5}FeO_{3}} \ch{Fe}-sublattice orders antiferromagnetically at high temperatures much above 400 K and $J^{\ch{Fe}-\ch{Fe}}$ interactions mainly account for the AFM in it \cite{chakraborty2018two}. $RE$-sublattice ordering in \ch{HoFeO_{3}} and \ch{DyFeO_{3}} systems has been reported to be below 10 K \cite{li2019spin,hoogeboom2021magnetic}. Our present study on \ch{Ho_{0.5}Dy_{0.5}FeO_{3}} was carried out at temperatures higher than 10 K hence, $J^{\ch{Fe}-\ch{Fe}}$ are dominant interactions and discussed in details. Distortions in the ideal perovskite unit cells result in distinct exchange interaction constants for the two nearest-neighbour \ch{Fe} spin-interactions; in-plane ($\mathit{ab}$-plane), $J_{\mathit{ab}}$ and between the  planes ($\vtr{c}$-direction), $J_{\mathit{c}}$ (See Fig.\ref{fig:1sch}(b)). Similarly, the next-nearest-neighbour exchange interaction between \ch{Fe} spins is crucial \cite{park2018low} and is represented by the constant, $J_{\mathit{NN}}$ (See Fig.\ref{fig:1sch}(b)).

Co-operative distortions in the unit cell result in displacement of oxygen ions and further lead to Dzyaloshinskii–Moriya interactions between the respective \ch{Fe} spins. The DM interactions coupled with exchange interactions manifest in canting of \ch{Fe} spins. Various DM interactions are denoted by respective DM vectors ($\vtr{D}_{14}$, $\vtr{D'}_{14}$, $\vtr{D}_{23}$, $\vtr{D'}_{23}$, $\vtr{D}_{34}$ and $\vtr{D}_{12}$) with green arrows in Fig.\ref{fig:1sch}(c) \cite{park2018low}.  

Space group \textit{P}bnm can have various representations such as $\Gamma_{1}$, $\Gamma_{2}$, $\Gamma_{3}$ and $\Gamma_{4}$, based on the manner in which different basis vectors of the spin configuration are aligned with respect to the lattice vectors, $\vtr{a}$, $\vtr{b}$ and $\vtr{c}$. Certain basis vectors for \ch{Fe} sublattice are defined in Eq.\ref{eq:BasisVectors} \cite{yamaguchi1974theory}.
\begin{subequations}
\label{eq:BasisVectors}
\begin{align}
2 \vtr{F} &= \vtr{S}_{1} + \vtr{S}_{2} + \vtr{S}_{3} + \vtr{S}_{4}\\
2 \vtr{G} &= \vtr{S}_{1} - \vtr{S}_{2} + \vtr{S}_{3} - \vtr{S}_{4}\\
2 \vtr{C} &= \vtr{S}_{1} + \vtr{S}_{2} - \vtr{S}_{3} - \vtr{S}_{4}\\
2 \vtr{A} &= \vtr{S}_{1} - \vtr{S}_{2} - \vtr{S}_{3} + \vtr{S}_{4}
\end{align}
\end{subequations}
The basis vector $\vtr{F}$ clearly represents respective net moment in the unit cell. Other basis vectors $\vtr{G}$, $\vtr{C}$ and $\vtr{A}$ represent the AFM behaviour of orthoferrites, where generally $\vtr{G}$ basis vector is the most prominent. Spin configurations of \ch{Fe} sublattice and their representations, $\Gamma_{1}$, $\Gamma_{2}$, $\Gamma_{3}$ and $\Gamma_{4}$ are shown in Fig.\ref{fig:1sch}(d), (e), (f) and (g), respectively. Notably, these three magnetic phases are represented in Bertaut's notations such as $\Gamma_{1}(A_{x}, G_{y}, C_{z})$, $\Gamma_{2}(F_{x}, C_{y}, G_{z})$, $\Gamma_{3}(C_{x}, F_{y}, A_{z})$ and $\Gamma_{4}(G_{x}, A_{y}, F_{z})$ \cite{bertaut1968representation}. In orthoferrites, typically at high temperatures $\Gamma_{4}(G_{x}, A_{y}, F_{z})$ magnetic phase is realised \cite{li2019spin} wherein, $\vtr{G}$, $\vtr{A}$ and $\vtr{F}$ are aligned along $\vtr{a}$, $\vtr{b}$ and $\vtr{c}$, respectively (See Fig.\ref{fig:1sch}(g)).

In 3D AFMs such as orthoferrites, phase transformations among $\Gamma_{1}(A_{x}, G_{y}, C_{z})$, $\Gamma_{2}(F_{x}, C_{y}, G_{z})$, $\Gamma_{3}(C_{x}, F_{y}, A_{z})$ and $\Gamma_{4}(G_{x}, A_{y}, F_{z})$ phases are commonly observed under the influence of temperature or magnetic field. Such transformations involve continuous (second-order transition) or abrupt (first-order transition) rotation of certain basis vectors and hence, they are popularly known as spin reorientation transitions. In the present case of \ch{Ho_{0.5}Dy_{0.5}FeO_{3}}, AFM order sets at high temperature well above 400 K, and  $\Gamma_{4}(G_{x}, A_{y}, F_{z})$ phase stabilizes. Even as the temperature is lowered, the $\Gamma_{4}(G_{x}, A_{y}, F_{z})$ phase persists down to $T_{SR1}$ = 49 K and then abruptly transforms to $\Gamma_{1}(A_{x}, G_{y}, C_{z})$ phase \cite{chakraborty2018two}. Further, around $T_{SR2}$ = 26 K, another transformation to $\Gamma_{2}(F_{x},C_{y},G_{z})$ phase has been reported \cite{chakraborty2018two}. Notably, under the application of magnetic field along $\mathit{c}$-axis, the $\Gamma_{1}(A_{x},G_{y},C_{z})$ phase was seen to transform to $\Gamma_{4}(G_{x}, A_{y}, F_{z})$ phase. The critical field required for this first-order transition increases with decreasing temperature \cite{chakraborty2018two}. It is apt to compare these phases with those in the parent orthoferrites, \ch{HoFeO_{3}} and \ch{DyFeO_{3}} \cite{li2019spin, mohammed2019magnetic, ovsyanikov2020neutron}. In \ch{HoFeO_{3}}, phase transitions occur in the following order as the temperature decreases: $\Gamma_{4} \xrightarrow{T_{1}} \Gamma_{412} \xrightarrow{T_{2}} \Gamma_{2} \xrightarrow{T_{3}}$ antiferromagnetically ordered \ch{Ho} phase, while in \ch{DyFeO_{3}} the following sequence is reported: $\Gamma_{4} \xrightarrow{T_{1}} \Gamma_{1} \xrightarrow{T_{2}} \Gamma_{5}$ (antiferromagnetically ordered \ch{Dy} phase). A recent report with the help of spin transport studies, has shown that \ch{Dy} spin moments get ordered due to the exchange interactions with \ch{Fe} spins in the temperature range, $T^{Dy}_{N}$ to 23 K \citep{hoogeboom2021magnetic}. We emphasize that the magnetic phase diagram of \ch{Ho_{0.5}Dy_{0.5}FeO_{3}} at low temperature is rich due to spin reorientation transitions, the related hystereses and a complex interplay between \ch{Fe} and $RE$ sublattices.

\begin{figure}\label{fig:Device}
\includegraphics[width=0.48\textwidth,keepaspectratio]{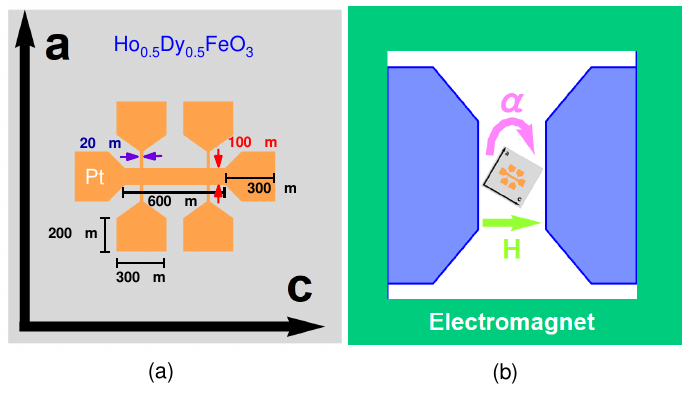}
\caption{Schematic of spin Hall magnetoresistance measurement: (a) Pt-Hall bar patterned on ac-plate of single crystal of \ch{Ho_{0.5}Dy_{0.5}FeO_{3}}, (b) SMR device mounted on a rotating probe inside electromagnet (Top view). Uniform constant magnetic field is applied in the plane of the device.
}
\end{figure}

\section{EXPERIMENTAL DETAILS}

Single crystal of \ch{Ho_{0.5}Dy_{0.5}FeO_{3}} was grown by optical floating-zone method \cite{chakraborty2018two}. The crystal was oriented using back-reflection Laue technique and cut in to a rectangular shaped $\mathit{ac}$-plate. The plate was polished with diamond paste (particle size $\approx0.25 $ $\mu m$) and the roughness of the surface was estimated using AFM as $R_{r.m.s.} \approx$ 2 nm. A Platinum (Pt) layer of 5 nm thickness was deposited on the polished surface of $\mathit{ac}$-plate by electron beam evaporation. Next, a Pt Hall bar was patterned using optical lithography which was followed by Argon ion beam etching (Refer \fig{fig:Device}(a) for dimensions and orientation of the Hall bar). For the RT SMR measurements, the sample was mounted on a custom-built rotating probe, which is incorporated with a Lakeshore electromagnet (2 T) (See \fig{fig:Device}(b)). Angular dependence of SMR was measured while rotating the sample in clock-wise ($\alpha$ scan) and counter clock-wise ($-\alpha$ scan) manner at various constant $H$ values in the range, 350 Oe to 16 kOe. Here, an ac current (333 Hz) of amplitude, $I_{\text{P-P}}$ $=$ 2 mA, was applied using Keithley 6221 DC/AC Current Source and the transverse voltage signal was measured with Stanford Research SR830 lock-in amplifier. For low-temperature SMR studies (down to 11 K), the device was mounted on a modified closed-cycle refrigerator equipped with rotating magnet arrangement (2400 Oe) and the transverse voltage was measured by passing $I_{\text{P-P}}$ $=$ 1 mA.

\section{Model of Spin Hall Magnetoresistance}

This section describes a simple model to compute SMR in \ch{Ho_{0.5}Dy_{0.5}FeO_{3}} single crystal. In order to compute SMR, we first estimate the equilibrium magnetic moment configuration of the magnetic unit cell. At high temperatures, \ch{Ho_{0.5}Dy_{0.5}FeO_{3}} exhibits $\Gamma_{4}$ magnetic phase in which the interactions between the \ch{Fe} spins are dominant. It is to be noted that $RE$ ordering occurs below 10 K and $RE$-\ch{Fe} interactions have been observed to persist and affect spin transport in \ch{DyFeO_{3}} system up to 23 K \cite{hoogeboom2021magnetic}. Therefore, we make a simple assumption that $RE$ spin-ordering is absent in the high temperature $\Gamma_{4}$ phase. However, paramagnetic moment of $RE$-sublattice can contribute to the total magnetization of the unit cell as it is influenced by both external magnetic field ($\Hext$) and local fields induced due to the ordering of the \ch{Fe}-sublattice.

We first define \ch{Fe} spin moment ($\vtr{S}_{\mathit{i}}$) and $RE$ spin moment ($\vtr{S}_{\mathit{k}}$) in Cartesian co-ordinate system (See Eqn.\ref{eq:Spins}(a) and (b)). The magnitude of \ch{Fe} spin moments is denoted by $S_{\ch{Fe}} = 5/2$ and that of $RE$ spin moments is $\left|\langle\mupara\rangle\right|$ i.e. statistical average of effective paramagnetic moment, $\mupara$.

\begin{widetext}

\begin{subequations}
\label{eq:Spins}

\begin{equation}
\vtr{S}_{\mathit{i}}=  S_{\mathit{i}}^{\mathit{x}} \hat{x} +  S_{\mathit{i}}^{\mathit{y}} \hat{y} +  S_{\mathit{i}}^{\mathit{z}} \hat{z}  \text{ such that } S_{\text{Fe}} = \sqrt{\left(S_{\mathit{i}}^{\mathit{x}}\right){}^2+\left(S_{\mathit{i}}^{\mathit{y}}\right){}^2+\left(S_{\mathit{i}}^{\mathit{z}}\right){}^2} \text{ and } \mathit{i} = 1, 2, 3, 4 
\label{eq:FeSpins}
\end{equation}

\begin{equation}
\vtr{S}_{\mathit{k}}= S_{\mathit{k}}^{\mathit{x}} \hat{x} +  S_{\mathit{k}}^{\mathit{y}} \hat{y} +  S_{\mathit{k}}^{\mathit{z}} \hat{z} \text{ such that } \left|\langle\mupara\rangle\right| = \sqrt{\left(S_{\mathit{k}}^{\mathit{x}}\right){}^2+\left(S_{\mathit{k}}^{\mathit{y}}\right){}^2+\left(S_{\mathit{k}}^{\mathit{z}}\right){}^2} \text{ and } \mathit{k} = 5, 6, 7, 8
\label{eq:RESpins}
\end{equation}
\end{subequations}

\begin{subequations}
\label{eq:Hamiltonian}
\begin{eqnarray}
\begin{aligned}
\mathcal{H}= &J_c\underset{\langle\mathit{i}\mathit{j}\rangle}{\overset{\text{Fe}}{\sum }}\vtr{S}_{\mathit{i}}\cdot \vtr{S}_{\mathit{j}}+J_{\text{ab}}\underset{\langle\mathit{i}\mathit{j}\rangle}{\overset{\text{Fe}}{\sum }}\vtr{S}_{\mathit{i}}\cdot \vtr{S}_{\mathit{j}}+J_{\text{NN}}\underset{\langle\langle\mathit{i}\mathit{j}\rangle\rangle}{\overset{\text{Fe}}{\sum }}\vtr{S}_{\mathit{i}}\cdot \vtr{S}_{\mathit{j}}+\underset{\langle\mathit{i}\mathit{j}\rangle}{\overset{\text{Fe}}{\sum }}\text{\textit{$\vtr{D}$}}_{\mathit{i}\mathit{j}}\cdot \left(\vtr{S}_{\mathit{i}}\times \vtr{S}_{\mathit{j}}\right)+K_a\overset{\text{Fe}}{\sum _{\mathit{i}} }\left(S_{\mathit{i}}^{\mathit{x}}\right){}^2\\&+K_c\overset{\text{Fe}}{\sum _{\mathit{i}} }\left(S_{\mathit{i}}^{\mathit{z}}\right){}^2-\overset{\text{Fe}}{\sum _{\mathit{i}} }\vtr{B}\cdot \vtr{S}_{\mathit{i}}
\end{aligned}
\label{eq:H1}
\end{eqnarray}

\begin{eqnarray}
\begin{aligned}
\mathcal{H}= &2J_c\left(\vtr{S}_1\cdot \vtr{S}_2+\vtr{S}_3\cdot \vtr{S}_4\right)+4J_{\text{ab}}\left(\vtr{S}_1\cdot \vtr{S}_4+\vtr{S}_2\cdot \vtr{S}_3\right)+8J_{\text{NN}}\left(\vtr{S}_1\cdot \vtr{S}_3+\vtr{S}_2\cdot \vtr{S}_4\right)+4J_{\text{NN}}\Bigl(\vtr{S}_1\cdot \vtr{S}_1\\
&+\vtr{S}_2\cdot \vtr{S}_2 {}+\vtr{S}_3\cdot \vtr{S}_3+\vtr{S}_4\cdot \vtr{S}_4\Bigr)+2\text{\textit{$\vtr{D}$}}_{14}\cdot \left(\vtr{S}_1\times \vtr{S}_4\right)+2\vtr{D'}_{14}\cdot \left(\vtr{S}_1\times \vtr{S}_4\right)+2\text{\textit{$\vtr{D}$}}_{23}\cdot \left(\vtr{S}_2\times \vtr{S}_3\right)\\
&+2\text{\textit{$\vtr{D'}$}}_{23}\cdot \left(\vtr{S}_2\times \vtr{S}_3\right)+2\text{\textit{$\vtr{D}$}}_{12}\cdot \left(\vtr{S}_1\times \vtr{S}_2\right)+2\text{\textit{$\vtr{D}$}}_{34}\cdot \left(\vtr{S}_3\times \vtr{S}_4\right)+K_a \left(\left(S_1^{\mathit{x}}\right){}^2+\Bigl(S_2^{\mathit{x}}\right){}^2{}+\left(S_3^{\mathit{x}}\right){}^2\\
&+\left(S_4^{\mathit{x}}\right){}^2\Bigr)+K_c \Bigl(\left(S_1^{\mathit{z}}\right){}^2+\left(S_2^{\mathit{z}}\right){}^2+\left(S_3^{\mathit{z}}\right){}^2+\left(S_4^{\mathit{z}}\right){}^2\Bigr)-\vtr{B}\cdot \left(\vtr{S}_1+\vtr{S}_2+\vtr{S}_3+\vtr{S}_4\right)
\end{aligned}
\label{eq:H2}
\end{eqnarray}

\end{subequations}

\end{widetext}

 Our simple Hamiltonian, $\mathcal{H}$ is described in Eqn.\ref{eq:Hamiltonian}(a), where the first two terms represent exchange interaction energy between the nearest-neighbour \ch{Fe} spins while the third term denotes exchange interaction between the next-nearest neighbour \ch{Fe} spins (See Fig.\ref{fig:1sch}(b)) as discussed in the Section \ref{Magnetism} on page \pageref{Magnetism}. Tilting of \ch{FeO_{6}} octahedra results in local shift of oxygen atoms and in-turn leads to DM interactions (fourth term in Eqn.\ref{eq:Hamiltonian}(a)). A local DM interaction associated with \ch{Fe}($i$)-\ch{O}-\ch{Fe}($j$) bond can be denoted by a vector, $\vtr{D}_{ij}$ which follows an anti-symmetric relation; $\vtr{D}_{ij} = -\vtr{D}_{ji}$  \cite{park2018low, mochizuki2009microscopic}. As depicted in Fig.\ref{fig:1sch}(c), one basal plane consists of \ch{Fe} spins, $\vtr{S}_{4}$ and $\vtr{S}_{1}$ while, the other plane comprises of $\vtr{S}_{3}$ and $\vtr{S}_{2}$. DM interactions in the former plane are described by the two different DM vectors, $\vtr{D}_{14}(\alpha_{ab},\beta_{ab},-\gamma_{ab})$ and $\vtr{D'}_{14}(-\alpha_{ab},\beta_{ab},-\gamma_{ab})$, while interactions in the latter plane are given by $\vtr{D}_{23}(\alpha_{ab},-\beta_{ab},-\gamma_{ab})$ and $\vtr{D'}_{23}(-\alpha_{ab},-\beta_{ab},-\gamma_{ab})$. Similarly, inter-plane DM interactions are denoted by $\vtr{D}_{34}(-\alpha_{c},\beta_{c},0)$ and $\vtr{D}_{12}(\alpha_{c},\beta_{c},0)$. In  order to realise $\Gamma_{4}(G_{x},A_{y},F_{z})$ phase in the absence of magnetic field, anisotropy energy is represented by fifth and sixth terms of Eqn.\ref{eq:Hamiltonian}(a), characterized by anisotropy constants, $K_{a}$ and $K_{c}$, respectively (such that $K_{a}$, $K_{c} < 0$). The last term in  Eqn.\ref{eq:Hamiltonian}(a) represents the Zeeman energy for \ch{Fe} moments. The magnetic flux density (magnetic induction) $\vtr{B}$ is defined as,
 \begin{equation}
\vtr{B} = \mu_0 (\Hext + \Mexp).
\label{eq:B1}
\end{equation}
Therefore, magnitude and direction of $\vtr{B}$ can be estimated with the applied magnetic field, $\Hext$ and experimentally measured respective magnetization, $\Mexp$. 

Eqn.\ref{eq:Hamiltonian}(a) can be rewritten as Eqn.\ref{eq:Hamiltonian}(b) after considering all relevant interactions within the nearest neighbour and the next-nearest neighbour \ch{Fe} spins. The values of all the constants ($J_{\mathit{c}}$, $J_{\mathit{ab}}$, $J_{NN}$, $\alpha_{\mathit{ab}}$, $\beta_{\mathit{ab}}$, $\gamma_{\mathit{ab}}$, $\alpha_{\mathit{c}}$, $\beta_{\mathit{c}}$, $K_{\mathit{a}}$, $K_{\mathit{c}}$ and $\mu_{\text{Para}}$) used in $\mathcal{H}$ (see Eqn.\ref{eq:Hamiltonian}(b)) are listed in Table \ref{tab:table1}.

\begin{center}
\begin{table}\centering
\begin{tabular}{ |c|P{1.5cm}|P{3cm}| }
\hline
\multirow{3}{10em}{Exchange Constants  (meV)} & $J_{\mathit{c}}$ & 5.2\\
&$J_{\mathit{ab}}$ &5.1\\
&$J_{NN}$ & 0.2\\ 
\hline
\multirow{5}{10em}{DM Vector Components (meV)} & $\alpha_{\mathit{ab}}$ & 0.0312\\
 &$\beta_{\mathit{ab}}$ &0.0294\\
&$\gamma_{\mathit{ab}}$ &0.0424\\
&$\alpha_{\mathit{c}}$ &0.0451\\
&$\beta_{\mathit{c}}$&0.1222\\
\hline
\multirow{2}{10em}{Anisotropy Constants (meV)} & $K_{\mathit{a}}$ & -0.0055\\
&$K_{\mathit{c}}$ & -0.00305\\
\hline
\multicolumn{2}{|c|}{\multirow{1}{*}{$\mu_{\text{Para}}$}} & 10.4\\
\hline

\end{tabular}
\caption{\label{tab:table1}Various constants and parameters used in the effective Hamiltonian for $\Gamma_{4}$ phase ($H = 0$ Oe) in \ch{Ho_{0.5}Dy_{0.5}FeO_{3}} system.}
\end{table}
\end{center}

Equilibrium spin configuration of \ch{Fe}-sublattice can be estimated by minimizing $\mathcal{H}$ with respect to $S_{\mathit{i}}^{\mathit{x}}$, $S_{\mathit{i}}^{\mathit{y}}$, $S_{\mathit{i}}^{\mathit{z}}$  ($\mathit{i} = 1$ to $4$). Subsequently, $RE$ moments $S_{\mathit{k}}$ ($\mathit{k} = 5$ to $8$) are determined considering their paramagnetic alignment (at high temperature) under the combined influence of $\Hext$ and the net moment of \ch{Fe}-sublattice ($\vtr{F}$), i.e. $\mu_0 (\Hext + \vtr{F})$.

In \ch{Ho_{0.5}Dy_{0.5}FeO_{3}} single crystal, we are studying angular dependence of transverse SMR by rotating $\Hext$ in $\mathit{ac}$-plane. We employ a simple model of SMR for ferromagnets that is described in the literature \cite{chen2013theory, nakayama2013spin}. This model has been used in a variety of complex magnetic phases such as collinear, canted and spiral AFMs because the magnetic unit cell is typically much smaller than the spin diffusion length of electrons in the adjacent platinum layer. According to this model, the transverse resistivity ($\rhot$) for applied $\Hext$ in the $\mathit{ac}$-plane can be given as,
\begin{eqnarray}
\label{Resistivity}
\rhot=\Delta \rho _1\langle S_{\mathit{i}}^{\mathit{x}} S_{\mathit{i}}^{\mathit{z}}\rangle + \Delta \rho _2 S_{\mathit{i}}^{\mathit{y}} + \Delta \rho _{\text{Hall}} H_{\mathit{b}}
\nonumber\\
\text{where, } \mathit{i} = 1 \text{ to } 8
\end{eqnarray}
Here, the first term is transverse SMR, the second term represents spin Hall induced anomalous Hall effect (SHAHE) and the last one is due to  the ordinary Hall effect (OHE). $\Delta \rho _1$ and $\Delta \rho _2$ are resistivities that are proportional to the real part ($G_{r}$) and the imaginary part ($G_{i}$) of the spin-mixing conductance, respectively. Notably, the SHAHE term depends on out-of-plane components (along $\mathit{b}$-axis) of the spin moments i.e. $S_{\mathit{i}}^{\mathit{y}}$. The ordinary Hall effect is governed by the Hall resistivity,  $\Delta \rho _{\text{Hall}}$ and out-of-plane component of the $\Hext$ i.e. $H_{b}$.

\section{Results and Discussion}

\subsection{Room Temperature Magnetization}
\label{RTMH}

\begin{figure}
\includegraphics[width=0.5\textwidth,keepaspectratio]{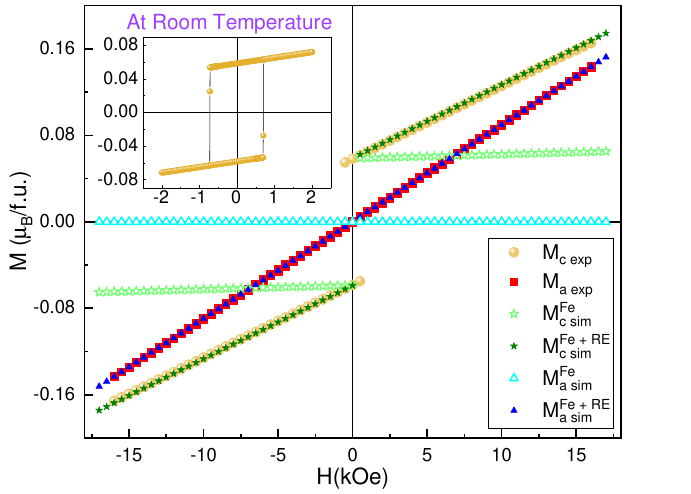}
\caption{\label{fig:MH} Isothermal magnetization (M) Vs. magnetic field (H) curves measured along $\vtr{a}$ and $\vtr{c}$ at 300 K. The inset shows magnified view of hysteresis in M-H curve along $\vtr{c}$.}
\end{figure}

In the absence of $\Hext$ at RT, $\Gamma_{4}$ phase is the stable phase in \ch{Ho_{0.5}Dy_{0.5}FeO_{3}} \cite{chakraborty2018two}. Figure \ref{fig:MH} shows magnetization ($M$) - field ($H$) isotherms $M_{\mathit{\text{c Exp}}}$ and $M_{\mathit{\text{a exp}}}$, measured while scanning the $\Hext$ in the range, -17 to 17 kOe at RT along $\mathit{c}$ and $\mathit{a}$-axis, respectively. The inset shows a magnified view of $M_{\mathit{\text{c exp}}}$ Vs. $\Hext$ curve highlighting a sharp switching accompanied by a hysteresis. The critical field ($H_{\text{c}}$) required for the switching between $\Gamma_{4}(+G_{x}, +F_{z})$ and $\Gamma_{4}(-G_{x}, -F_{z})$ domains is estimated to be $\approx$ 713 Oe. In order to understand experimentally-measured $M$-$H$ curves along $\mathit{c}$ and $\mathit{a}$-axis at RT, we simulated the equilibrium magnetization by minimizing $\mathcal{H}$ in Eqn.\ref{eq:Hamiltonian}(b). We realized that the simulated magnetization considering \ch{Fe} sublattice alone $M^{\ch{Fe}}_{\mathit{\text{c sim}}}$ ($M^{\ch{Fe}}_{\mathit{\text{a sim}}}$) can not account for the experimentally measured magnetizations, $M_{\mathit{\text{c exp}}}$ ($M_{\mathit{\text{a exp}}}$) as shown in Fig.\ref{fig:MH}. Accordingly, we consider that an additional contribution is required and that may come from the paramagnetic alignment of $RE$ spins, induced due to the ordering of \ch{Fe}-sublattice. Further, considering the combined effect of ordered \ch{Fe}-sublattice and paramagnetic $RE$-sublattice, we simulated the magnitude of magnetization, $M^{\ch{Fe}+RE}_{\mathit{\text{c sim}}}$ ($M^{\ch{Fe}+RE}_{\mathit{\text{a sim}}}$) and matched it to $M_{\mathit{\text{c exp}}}$ ($M_{\mathit{\text{a exp}}}$) as shown in \fig{fig:MH}.  

It is clear that the paramagnetic term of the $RE$-sublattice is essential; however, only a certain fraction i.e. 19.8 \% (27.6 \%) of this term was needed to match the observed $M_{\mathit{\text{c exp}}}$ ($M_{\mathit{\text{a exp}}}$) at RT while computing $M^{\ch{Fe}+RE}_{\mathit{\text{c sim}}}$ ($M^{\ch{Fe}+RE}_{\mathit{\text{a sim}}}$). The details hint at the posiibility of weak AFM ordering between \ch{Fe} and $RE$-sublattice. A more detailed analysis is required to substantiate this inference.

\subsection{Room Temperature SMR studies}
\label{RTSMR}

\begin{figure*}
\includegraphics[width=\textwidth,keepaspectratio]{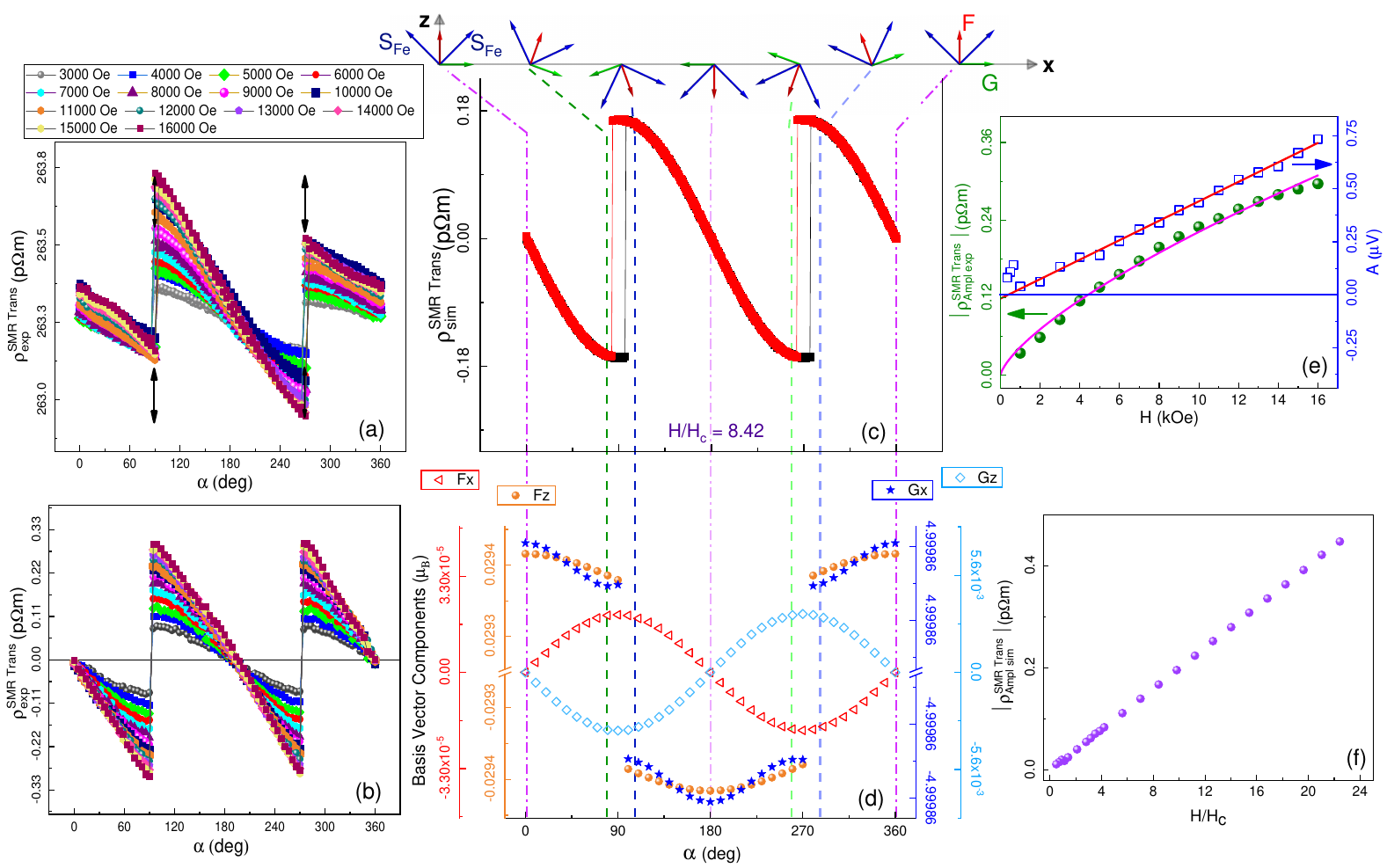}
\caption{\label{fig:SMRHighH} Transverse SMR measured for $\alpha$ and -$\alpha$ scans at RT under application of various constant in-plane magnetic fields in the range of 3 kOe to 16 kOe. (a) Angular dependence of raw SMR data with asymmetry marked by black arrows, (b) Corrected SMR vs. $\alpha$ after removal of sinusoidal contribution, Angular variation of equilibrium spin configuration ($\Gamma_{4}$ phase) in ac-plane is computed; (c) Simulated SMR data, (d) Respective basis vector components (Fx, Fz, Gx and Gz) vs. $\alpha$, (e) Field variation of extracted amplitude of sinusoidal contribution (blue squares) from raw SMR data and the amplitude of corrected SMR data (green spheres), (f)  Amplitude of simulated SMR as a function of magnetic field.}
\end{figure*}

In orthoferrites, magnetic phase is influenced by both temperature and applied magnetic field. Under the application of high fields along $\vtr{a}$, $\vtr{b}$ and $\vtr{c}$, the spin moment configuration tends to stabilize in $\Gamma_{2}$, $\Gamma_{3}$ and $\Gamma_{4}$ phases, respectively {(see \fig{fig:1sch})}. In our spin transport studies, $\Hext$ of constant magnitude is rotated in the ${\mathit{ac}}$-plane of \ch{Ho_{0.5}Dy_{0.5}FeO_{3}} single crystal. Angular dependence of $\rhote$ was measured at different specific $H$ values between 350 Oe to 16 kOe. $\Hext$ was rotated in both forward and reverse sense and the respective $\rhote$ -scans, $\alpha$ scan and -$\alpha$ scan were recorded. We classify our SMR studies in to two field regimes; 1) $H > H_{\text{c}}$, and 2) $H < H_{\text{c}}$.

\subsubsection{$H > H_{\text{c}}$}
\label{RTSMRaboveHc}

We will first discuss the SMR results at high field (3 to 16 kOe) well above the $H_{\text{c}}$ (713 Oe) (See \Fig{fig:SMRHighH}(a)). $\rhote$ varies gradually and exhibits sudden changes around $\alpha$ = 90 deg and 270 deg. A narrow hysteresis ($4.5\text{ deg}$) is seen in $\rhote$ which matches the step-size of the $\alpha$ scan i.e. $\Delta \alpha$. The hysteresis, which comprises of step-like sharp change in $\alpha$ and $-\alpha$ scans, is centred around $\vtr{a}$ ($-\vtr{a}$) i.e. $\alpha$ = 90 deg (270 deg). However, the data can not be explained solely on the basis of expected periodicity of 180 deg in SMR as marked by arrows in the figure. It is prudent to examine the possible contributions in the overall signal that originates from SHAHE and OHE having periodicity of 360 deg (refer to Eqn.\ref{Resistivity}). In order to estimate SMR contribution in $\rhote$, we have removed $A\sin(\alpha+\delta)$ magnitude from the experimental curve where $A$ and $\delta$ are the respective amplitude and phase difference of SHAHE or OHE contribution. The separated contributions of $\rhote$ and $A\sin(\alpha+\delta)$ signals are characterized by $\rho_{\text{Ampl exp}}^{\text{SMR Trans}}$ and $A$, respectively and are plotted as a function of $H$ in Fig.\ref{fig:SMRHighH}(e). The field variation of $\rho_{\text{Trans}}^{\text{SMR Amplitude}}$ is found to be non-linear which is fitted with a function, $\text{c} H^{\gamma}$. The fit-parameters c and $\gamma$ are estimated as 3.31$\times 10^{-16}$ and 0.71, respectively. On the other hand, the overall field variation of $A$ is linear with a sharp discontinuity at low field ($\approx$ 650 Oe). The high-field ($H > H_{\text{c}}$) contribution extracted from modulated $\rhote$ can not be explained by SHAHE as its amplitude increases with increase in $H$. On the other hand, OHE caused by out-of-plane (OOP) tilting of the sample is a likely mechanism to explain the sinusoidal contribution as extracted from the modulation of $\rhote$. For instance, if $H_{ext} =$ 16 kOe is applied along $\mathit{a}$-axis, the sample needs to be tilted about $\mathit{c}$-axis just by $\approx$ 0.17 deg to match the extracted amplitude of the sinusoidal signal i.e. $A =$ 0.734 $\upmu \text{V}$ (Ordinary Hall-coefficient for 6.5 nm thick Platinum is $\approx$ 23.1 p$\Omega$m/T \cite{meyer2015anomalous}). 

After removal of the OHE contribution, the corrected $\rhote$ Vs. $\alpha$ curves are plotted in \Fig{fig:SMRHighH}(b). The periodicity of the curves is 180 deg as expected for the typical SMR modulation. Angular variation of $\rhote$ shows continuous behaviour except in the vicinity of $\vtr{a}$ ($\alpha = 90\text{ deg}$) and $\vtr{-a}$ ($\alpha = 270\text{ deg}$). Notably, the value $\rhote$ is zero near $\vtr{c}$ ($\alpha = 0\text{ deg}$) and $\vtr{-c}$ ($\alpha = 180\text{ deg}$). Our $\rhote$ data exhibits negative SMR. It has been shown that AFM N\'eel (basis) vector $G$ and FM basis vector $F$ yields negative and positive SMR, respectively \cite{fischer2018spin, geprags2020spin}. The observed negative SMR in \ch{Ho_{0.5}Dy_{0.5}FeO_{3}} may be due to the dominance of $G$ over $F$.  

In order to understand the modulation of $\rhote$ and the underlying mechanism, we first minimized $\mathcal{H}$ in Eqn.\ref{eq:Hamiltonian} using parameters listed in Table \ref{tab:table1}, determined equilibrium spin configuration, included paramagnetic contribution of RE-sublattice and then, computed $\rhot$ using Eqn.\ref{Resistivity}. As discussed earlier, the paramagnetic alignment of RE-sublattice has significant contribution in magnetization. However, in the \ch{Fe}-sublattice as $G$ dominates over $F$ in determining the SMR, the paramagnetic contribution to SMR is found negligible compared to that in the \ch{Fe}-sublattice. We note that all exchange, DM and anisotropy terms are essential in the Hamiltonian and affect the equilibrium spin configuration and computed values of SMR. As discussed in Eqn.\ref{eq:B1}, $\vtr{B}$ in the Zeeman energy term was calculated using the experimentally measured magnetization $M_{\mathit{\text{c exp}}}$ and $M_{\mathit{\text{a exp}}}$ along $\vtr{c}$ and $\vtr{a}$, respectively. This simple approach helped to simulate the hysteresis associated with $\alpha$ and -$\alpha$ scans which resembles our experimental result. The simulated $\rhot$ ($\alpha$ and -$\alpha$ scans) at $H/H_{\text{c}}$ = 22.5 is shown in \Fig{fig:SMRHighH}(c). A comparison of $\rhot$ with $\rhote$ data yielded the value of $\Delta\rho_{1}$ to be $0.8\times10^{-10}$ $\Omega m$. A schematic describing angular evolution of the spin configuration is represented at the top of the figure. Figure \ref{fig:SMRHighH}(d) shows angular variation of basis vector components ($F_{x}$, $F_{z}$, $G_{x}$ and $G_{z}$) of the spin configuration. Angular modulation of $F_{x}$ and $G_{z}$ is continuous while, $F_{z}$ and $G_{x}$ show sharp changes around $\alpha = 90\text{ deg}$ and $\alpha = 270\text{ deg}$, pointing towards sudden switching between $\Gamma_{4}(G_{x},F_{z})$ and $\Gamma_{4}(-G_{x},-F_{z})$ domains. We argue that the switching and related rotational hysteresis are governed by $H_{c}$. In the $\alpha$ scan, both $F$ and $G$ vectors are confined and rotate in ${\mathit{ac}}$-plane, where they remain perpendicular to each other. 

The variation of calculated $\rho^{\text{SMR Trans}}_{\text{sim Amp}}$ Vs. $H/H_{c}$ shows a linear behaviour (See \Fig{fig:SMRHighH}(f)). The change in SMR-amplitude can be understood in the following manner; In the absence of anisotropy, during $\alpha$-scan the $\Gamma_{4}(G_{x},F_{z})$ domain would rotate synchronously with rotating $\vtr{H}$ and result in -$\sin(2 \alpha)$ behaviour of SMR with extremum value at $45\text{ deg}$. On the other hand, in the presence of strong anisotropy, the $\Gamma_{4}(G_{x},F_{z})$ domain rotates to a small angle (yielding an increase in SMR), before it suddenly switches to the $\Gamma_{4}(-G_{x},-F_{z})$ domain (reversal in the sign of the SMR). As $\vtr{H}$ increases, the Zeeman energy starts to dominate the anisotropy and the $\Gamma_{4}(G_{x},F_{z})$ domain rotates to higher angle before switching to $\Gamma_{4}(-G_{x},-F_{z})$. This is a possible cause for the observed increase in the amplitude of SMR with increase in $H$. Similarly, for a fixed ${H}$-value, if the anisotropy is weaken then, the SMR amplitude would increase. 

\begin{figure}
\includegraphics[width=0.48\textwidth,keepaspectratio]{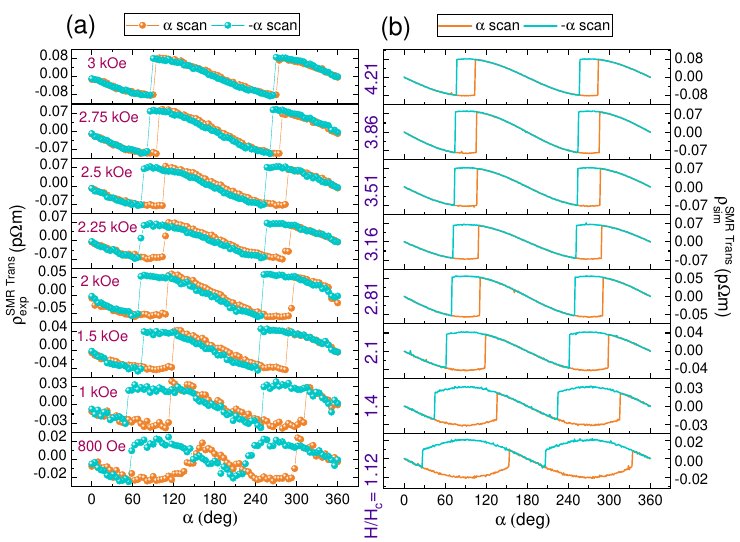}
\caption{\label{fig:SMRaboveHc}Transverse SMR ($\alpha$ and -$\alpha$ scans) curves are shown for various magnetic field values just above a critical field ($H_{c}$) and are denoted by corresponding normalized magnetic fields (H/$H_{c}$ = 1.12, 1.4, 2.1, 2.81, 3.16, 3.51, 3.86 and 4.21); (a) Experimentally measured SMR data (at room temperature) for various fields (range: 800 Oe to 3 kOe) above a critical field of Hc = 713 Oe. (b) Simulated SMR data at various corresponding H/Hc values}
\end{figure}

In order to probe mechanism behind the observed hysteresis in $\rhote$ curves, we carried out SMR measurements at various fields (from 800 Oe to 3 kOe) just above $H_{\text{c}}$ (713 Oe) as shown in \Fig{fig:SMRaboveHc}(a). $\rhote$ Vs. $\alpha$ curves are highly skewed and deviate substantially from $\sin(2 \alpha)$ behaviour (typically observed in collinear magnets at saturation fields) indicating strong anisotropy. At 800 Oe, $\alpha$ and -$\alpha$ scans of $\rhote$ exhibit sharp changes accompanied by broad rotational hysteresis near $\alpha$ = 90 deg and 270 deg (Fig.\ref{fig:SMRaboveHc}(a)). It is discernible that the hysteretic region decreases in size with increase in $H$. The hysteresis observed in ${\mathit{ac}}$-scan on \ch{Ho_{0.5}Dy_{0.5}FeO_{3}} could be interpreted as due to the first-order switching between $\Gamma_{4}(G_{x},F_{z})$ and $\Gamma_{4}(-G_{x},-F_{z})$ domains \cite{bazaliy2004spin}. For instance, in alpha scan, $H$ cos($ \alpha $) is the projection of $\Hext$ along $\vtr{c}$ i.e. $H_{c}$ . When $\alpha > 90\text{ deg}$, $\Hext$ becomes negative (i.e. along $-\vtr{c}$) and further, when it equals to $-H_{c}$ (i.e. coercivity of magnetization hysteresis along $\vtr{c}$), the switching occurs suddenly and manifests in the sign reversal of $\rhote$ (a step-like feature in the SMR). It is evident that for a given value of $-H_{c}$, with increase in magnitude of $H$ the width of the hysteresis decreases as seen in \fig{fig:SMRaboveHc}(a). To compare with $\rhote$ Vs. $\alpha$ curves, we simulated the $\rhot$ Vs. $\alpha$ curves for a corresponding set of normalized magnetic fields (H/$H_{c}$ = 1.12, 1.4, 2.1, 2.81, 3.16, 3.51, 3.86 and 4.21) as shown in Fig.\ref{fig:SMRaboveHc}(b). Reduction in hysteretic region with increase in $H$ indicates the competition of Zeeman energy with the anisotropy energy.

\subsubsection{$H < H_{\text{c}}$}
\label{RTSMRbelowHc}

\begin{figure}
\includegraphics[width=0.48\textwidth,keepaspectratio]{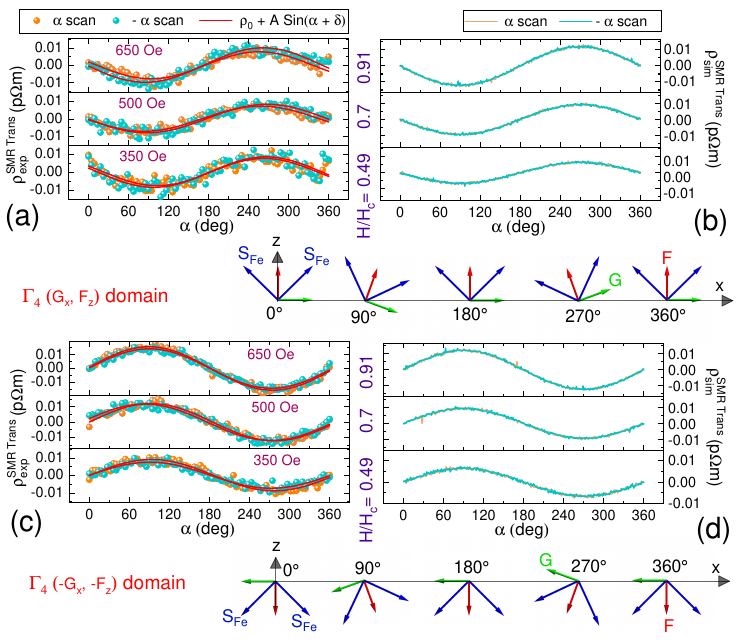}
\caption{\label{fig:smrRTlowH} Transverse SMR ($\alpha$ and -$\alpha$ scans) curves are plotted for various magnetic field values just below a critical field (Hc) and are denoted by corresponding normalized magnetic fields (H/Hc = 0.49, 0.7 and 0.91); (a) (or (c)) A domain $\Gamma_{4}(G_{x},F_{z})$ (or $\Gamma_{4}(-G_{x},-F_{z})$) was prepared by applying field of 2 kOe (or -2 kOe) along $\vtr{c}$ and then, SMR data (at room temperature) was experimentally recorded for fields (350, 500 and 650 Oe) below a critical field of Hc = 713 Oe. The data are fitted with sine function. (b and d) Simulated SMR data for corresponding H/Hc values.}
\end{figure}
  
In the absence of $\Hext$, $\Gamma_{4}$ phase possesses two possible degenerate spin configurations (domains); $\Gamma_{4}(G_{x},F_{z})$ and $\Gamma_{4}(-G_{x},-F_{z})$. Here, a single domain can be achieved by application of $H$ higher than the coercive field $H_{c}$ along $\vtr{c}$ or $-\vtr{c}$. In order to obtain a single domain of $\Gamma_{4}(G_{x},F_{z})$, we applied 2 kOe field along $\vtr{c}$ and subsequently, reduced the field to a value (i.e. 350, 500 or 650 Oe) below $H_{c}$ and then, carried out the $\alpha$-scan measurement. Interestingly, the angular dependence of $\rhote$ measured at $H =$ 350, 500 and 650 Oe yielded a sinusoidal modulation with periodicity of 360 deg as illustrated in Fig.\ref{fig:smrRTlowH}(a). Notably, the 360 deg-periodicity behaviour does not match with the characteristic 180 deg-periodicity of the SMR in collinear magnets. However, detailed studies are needed to understand the anomalous modulation of the observed signal. It is to be noticed that $\rhote$ comprises of three different mechanisms (see Eqn.\ref{Resistivity}). The first two mechanisms namely, SMR and SHAHE are explicitly related to the spin current absorption by in-plane and out-of-plane components of the spin moments at \ch{Ho_{0.5}Dy_{0.5}FeO_{3}}$\vert$\ch{Pt} interface, respectively. Whereas, the third mechanism is OHE that is solely dependent on OOP component of the $\Hext$.
 
In order to examine the possible role of OHE, we carried out $\rhote$ measurements at $H =$ 350, 500 and 650 Oe after obtaining another single domain $\Gamma_{4}(-G_{x},-F_{z})$ by employing 2 kOe field along $-\vtr{c}$ (Refer Fig.\ref{fig:smrRTlowH}(c)). It is discernible that the $\rhote$ modulations are reversed in sign (shifted by 180 deg) for all the field values. This result discards the possibility of OHE as an underlying mechanism. However, the contribution from SHAHE may arise due to the presence of OOP component of the spin moments and the switching of such moments can result in the modulation of the SHAHE signal with the periodicity of 360 deg. The curves in  Fig.\ref{fig:smrRTlowH}(a) and (c) are fitted with $\rho_{0} + A \sin{(\alpha + \delta)}$ where, $\rho_{0}$ is the offset resistivity originating from longitudinal component of the resistivity due to finite width of transverse leads, $A$ is the amplitude and $\delta$ the phase difference. The fitted amplitudes $A$ are plotted as a function of $H$ in Fig.\ref{fig:SMRHighH}(e) and are found to vary linearly with distinct slope compared to that for $H > H_{c}$. The linear increase in amplitude with $H$ is contrary to the general observation that high values of in-plane field tend to decrease the OOP component of the magnetization. Therefore, SMR could be a possible mechanism and it warrants further investigation. 

For the two different domains $\Gamma_{4}(G_{x},F_{z})$ and $\Gamma_{4}(-G_{x},-F_{z})$, we simulated $\rhot$ Vs. $\alpha$ curves for corresponding normalized magnetic fields just below the critical field (H/$H_{c}$ = 0.49, 0.7 and 0.91) as shown in Fig.\ref{fig:smrRTlowH}(b) and (d) and compared with $\rhote$ Vs. $\alpha$ curves in Fig.\ref{fig:smrRTlowH}(a) and (c), respectively. Besides, the angular evolution of spin configuration of $\Gamma_{4}(G_{x},F_{z})$ and $\Gamma_{4}(-G_{x},-F_{z})$ domains are represented by respective schematics in Fig.\ref{fig:smrRTlowH}). At low $H$ ($H < H_{\text{c}}$), as $H$ rotates in ${\mathit{ac}}$-plane, Zeeman energy is not sufficient to initiate switching between $\Gamma_{4}(G_{x},F_{z})$ and $\Gamma_{4}(-G_{x},-F_{z})$ domains. Here, in $\alpha$ scan (for either domain), both basis vectors $\vtr{F}$ and $\vtr{G}$ oscillate smoothly about their mean position with 360 deg periodicity. It manifests in sinusoidal smooth modulation of $\rhot$ with periodicity of 360 deg. Our $\rhot$ curves (Fig.\ref{fig:smrRTlowH}(b) and (d)) are in good agreement with $\rhote$ curves (Fig.\ref{fig:smrRTlowH}(a) and (c)). The simulations reveal that the SMR solely accounts for the observed 360 deg modulation of $\rhote$. Here, we would like to highlight that the 360 deg modulation of SMR at $H < H_{\text{c}}$, has potential in magnetic field-direction sensing devices. 

Finally, we revisit and discuss key points related to the choice of ${\mathit{ac}}$-plane for the SMR studies on \ch{Ho_{0.5}Dy_{0.5}FeO_{3}}$\vert$\ch{Pt} hybrid. Typically, in $\Gamma_{4}(G_{x}, A_{y}, F_{z})$ and $\Gamma_{2}(F_{x}, A_{y}, G_{z})$ phases, both $\vtr{F}$ and $\vtr{G}$ basis vectors are prominent where, $\vtr{G}$ tends to align perpendicular to $\Hext$ while, the $\vtr{F}$ favours parallel alignment with $\Hext$. Therefore, for our spin transport studies on $\Gamma_{4}(G_{x}, A_{y}, F_{z})$ phase at RT and investigation of the SRT to $\Gamma_{2}(F_{x}, A_{y}, G_{z})$ phase at lower temperature, we chose the ${\mathit{ac}}$-plane of \ch{Ho_{0.5}Dy_{0.5}FeO_{3}} single crystal. Interestingly, this choice enabled us to observe a rotational hysteresis in the SMR signal, which can be associated with the switching between $\Gamma_{4}(G_{x},F_{z})$ and $\Gamma_{4}(-G_{x},-F_{z})$ domains. Here, we would like to mention that in a recent study on \ch{DyFeO_{3}} single crystal, the SMR on ${\mathit{ab}}$-plate exhibited no such hysteresis \cite{hoogeboom2021magnetic}. We argue that this could be possibly due to an inaccessibility of the switching between $\Gamma_{4}(G_{x},F_{z})$ and $\Gamma_{4}(-G_{x},-F_{z})$ domains in their ${\mathit{ab}}$-plane scan-configuration. In \ch{Ho_{0.5}Dy_{0.5}FeO_{3}}$\vert$\ch{Pt} hybrid, both $\vtr{F}$ and $\vtr{G}$ basis vectors stay confined in the ${\mathit{ac}}$-plane during the $\alpha$-scan, manifesting in intriguing angular modulation of the SMR.

\subsection{Low Temperature Two-fold Spin Re-orientation Transition, $\Gamma_{4}$ $\rightarrow$ $\Gamma_{1}$ $\rightarrow$ $\Gamma_{2}$}
 
  \begin{figure}
\includegraphics[width=0.48\textwidth,keepaspectratio]{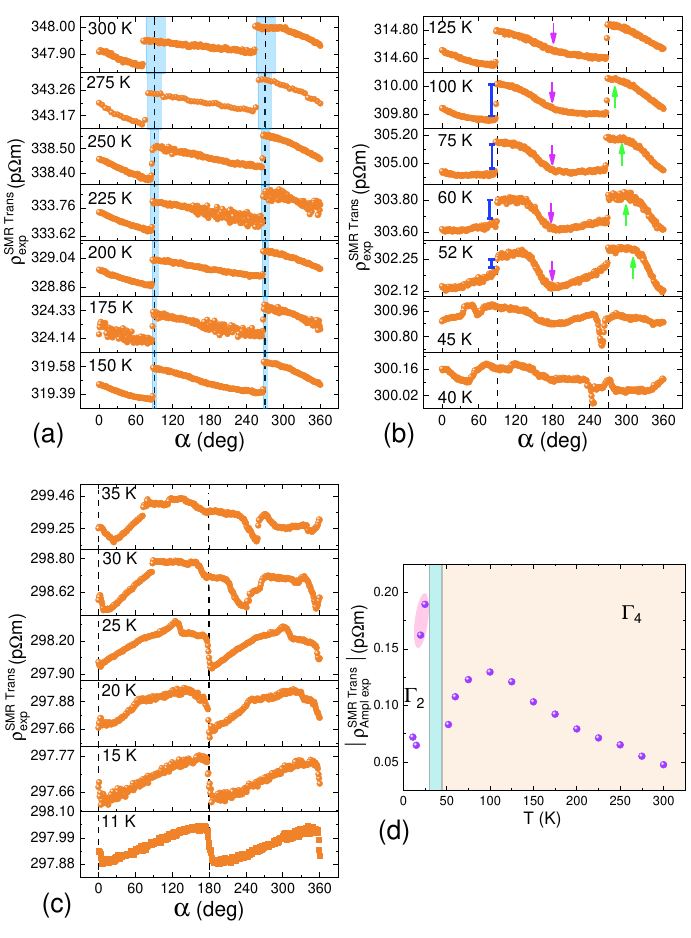}
\caption{\label{SMRLowT} Transverse SMR measured (2.4 kOe) at different temperatures in the range: (a) 300-150 K (An apparent rotational hysteresis is denoted by shaded blue region), (b) 125 to 40 K (The shift in peak is marked by green arrows, the bending of $\rhote$ near $\alpha=$180 deg is denoted using purple arrows and the jump in $\rhote$ near $\alpha$ = 270 deg is shown by blue mark), and (c) 35-11 K (Spin reorientation transition to $\Gamma_{2}$). (d) Temperature evolution of transverse SMR amplitude.}
\end{figure}

The angular dependence of $\rhote$ (-$\alpha$-scan) was determined by rotating $\Hext$ ($H$ = 2.4 kOe) in the $\mathit{ac}$-plane at various temperatures in the range, 11 to 300 K (See Fig.\ref{SMRLowT} (a) to (c)). In the absence of $\Hext$, the high-temperature $\Gamma_{4}$ phase in $\ch{Ho_{0.5}Dy_{0.5}FeO_{3}}$ persists down to 49 K before transforming to $\Gamma_{1}$ phase. As discussed in the previous section, the modulation of $\rhote$ in $\Gamma_{4}$ phase exhibits a sharp change accompanied by a rotational hysteresis while crossing $\vtr{a}$ (See Fig.\ref{fig:SMRHighH}). The $\alpha$ scans were not recorded at low temperatures; however, the hysteretic region could be identified by the shift in angle (overshoot below 90 deg or 270 deg) at which the sharp change occurred in $\rhote$. The width of the hysteresis, characterized by the overshoot-angle, is represented by a blue-shaded region and this width is found to decrease systematically with decrease in the temperature as shown in Fig.\ref{SMRLowT}(a). As discussed in Section \ref{RTSMR}, the origin of hysteresis in the $\alpha$-scan lies in the first-order switching of the domains ($\Gamma_{4}(G_{x},F_{z})$ $\leftrightarrow$ $\Gamma_{4}(-G_{x},-F_{z})$) where, the angle ($\alpha$) at which the switching occurs can be estimated by the relation; $H\cos\alpha= H_{c}$. Further, the width of the hysteresis is given by 2 $\times\mid(90\text{ deg}-\alpha)\mid$. Notably, for all -$\alpha$-scans recorded at low temperature, $H$ is constant ($=$ 2.4 kOe) and it leads to $H_{c}$ $\varpropto$ $\cos\alpha$. Hence, the observed decrease in the width of the hysteresis can be directly related to the reduction in the magnitude of $H_{c}$. This suggests weakening of the anisotropy and favours $\vtr{F}$ ($\vtr{G}$) along $\mathit{c}$-axis ($\mathit{a}$-axis). Until 125 K, the amplitude of modulation, $\rho^{\text{SMR Trans}}_{\text{exp Ampl}}$ in -$\alpha$-scan is determined by the magnitude of the sudden jump in $\rhote$ ($\rho^{\text{SMR Trans}}_{\text{exp Jump}}$, denoted by blue mark) around 90 deg and 270 deg  ($\rho_{\text{SMR Trans}}^{\text{exp Ampl}} = \rho^{\text{SMR Trans}}_{\text{exp Jump}}/2$) (Fig.\ref{SMRLowT}(a) and (b)). However, below 100 K it does not hold true as the value of $\rho^{\text{SMR Trans}}_{\text{exp Jump}}$ tends to vanish systematically at lower temperatures until 52 K. At $T \geq 125$ K, the $\rhote$ curve peaks around 90 deg (270 deg) and in the range of 100 to 52 K, the peak-position (shown by green arrow) shifts towards lower angles, possibly tending towards $\alpha=$ 45 deg (225 deg) (See Fig.\ref{SMRLowT}(b)). Besides, a systematic bending of $\rhote$ curve (marked by a purple arrow) is seen around $\alpha=$ 180 deg in the same $T$-interval. These evolutions involving $\rho^{\text{SMR Trans}}_{\text{exp Jump}}$, the peak shift and the bending, hint towards a gradual reduction in the skewness of $\rhote$-modulation tending possibly towards an ideal $\sin 2\alpha$ behaviour. This reduction in the skewness of $\rhote$ curve is related to a further weakening of the anisotropy. An earlier report claimed that the presence of \ch{Fe}-$RE$ interactions in orthoferrites, demands use of effective (modified) anisotropic constants \cite{nikitin2018decoupled}. As the strength of \ch{Fe}-$RE$ interactions varies with temperature \cite{yamaguchi1974theory}, the related effective anisotropic constants are temperature dependent. It is known that the strength of \ch{Fe}-$RE$ interactions increases with decrease in the temperature and such increase in the strength of interactions could be the main mechanism responsible for the observed weakening of the anisotropy (See \fig{SMRLowT}(a) and (b)).

In the temperature interval of 45 to 30 K, $H$-$T$ phase diagram is complex involving different phases ($\Gamma_{4}$, $\Gamma_{1}$ and $\Gamma_{2}$) and hysteretic regions consisting of co-existing metastable phases \citep{chakraborty2018two} (See \Fig{SMRLowT}(b) and (c)). $\rhote$ curves recorded at 45, 40, 35 and 30 K have multiple complex features. At temperatures below 25 K, a sharp step-like feature in $\rhote$ curve is seen to occur around $\vtr{c}$ indicating a dominance of anisotropy favours $\vtr{G}$ along $\mathit{c}$-axis in the $\Gamma_{2}$ phase (See Fig.\ref{SMRLowT}(c)). Thus, the SMR was found to be a useful tool to identify $\Gamma_{4}$ and $\Gamma_{2}$ phases and their respective anisotropies.

In order to estimate the amplitude of SMR-modulation ($\rho^{\text{SMR Trans}}_{\text{exp Ampl}}$) at a particular temperature, we have removed $A\sin(\alpha+\delta)$ contribution from $\rhote$ curve as discussed in the previous Section\ref{RTSMR}. The temperature evolution of $\rho^{\text{SMR Trans}}_{\text{exp Ampl}}$ is plotted in Fig.\ref{SMRLowT}(d). In the $\Gamma_{4}$ phase, $\rho^{\text{SMR Trans}}_{\text{exp Ampl}}$ is seen to increase as the temperature is lowered to 100 K. Below this temperature, $\rho^{\text{SMR Trans}}_{\text{exp Ampl}}$ decreases smoothly till 52 K. Earlier, we have argued that as ${T}$ is lowered to 100 K, the anisotropy is systematically weaken. This weakening of anisotropy compared to the Zeeman energy (of fixed value), can cause an increase in the $\mid\rho^{\text{SMR Trans}}_{\text{exp Ampl}}\mid$ value (Refer to the discussion in Subsection \ref{RTSMRaboveHc}). Since the $T$-interval, 45 to 30 K (represented by cyan colour), exhibits the complex features in the -$\alpha$-scan curves, we are unable to comment on the $\rho^{\text{SMR Trans}}_{\text{exp Ampl}}$ variation in this range. Within the $\Gamma_{2}$ phase ($T \leq$ 25 K), the variation of $\rho^{\text{SMR Trans}}_{\text{exp Ampl}}$ shows a drastic change between 15 and 20 K. The anomalously large values observed at 20 and 25 K are marked by pink-shaded region. The observed anomaly warrants further investigation.

\section{CONCLUSION}
In this work, we have employed spin Hall magnetoresistance (SMR) as a probe to examine the magnetic anisotropy and spin reorientation transition (SRT) in a complex 3D antiferromaget, \ch{Ho_{0.5}Dy_{0.5}FeO_{3}} possessing weak ferromagnetism (WFM).  At various constant temperatures (range: 300 to 11 K), SMR was measured on a  $\mathit{b}$-plate of the single crystal \ch{Ho_{0.5}Dy_{0.5}FeO_{3}} $\vert Pt$ heterostructure by passing a charge current along $\mathit{c}$-direction and rotating the magnetic field ($\Hext$) in the $\mathit{ac}$-plane ($\alpha$-scan). In the room-temperature $\Gamma_{4}$ phase the basis vectors, $\vtr{G}$ (N\'eel vector) and $\vtr{F}$ (WFM) lie in the $\mathit{ac}$-plane. Here two competing energies, the anisotropy energy (favouring $\vtr{G}\parallel \vtr{a}$ and $\vtr{F}\parallel \vtr{c}$) and the Zeeman energy (promoting $\vtr{F}\parallel \Hext$ and $\vtr{G}\perp \Hext$), play a key role in the observed SMR-modulation. In $\Gamma_{4}$ phase, a critical field $H_{c} = 713$ Oe was found required to overcome the anisotropy energy and cause switching between the two degenerate domains; $\Gamma_{4}(G_{x},F_{z})$ and $\Gamma_{4}(-G_{x},-F_{z})$. Transverse SMR ($\rhote$) curves ($\alpha$ and -$\alpha$ scans) recorded at $H > H_{c}$, exhibited a sharp change (a sign-reversal) accompanied by a rotational hysteresis near $\mathit{a}$-axis. The sign-reversal and associated hysteresis are due to the first-order switching between the two domains. The hysteretic region was found to become narrower as $H$ was increased. Further, a single domain of $\Gamma_{4}(G_{x},F_{z})$ ($\Gamma_{4}(-G_{x},-F_{z})$) was achieved using $H = 2$ ($-2$) kOe (here $H > H_{c}$) and then, the SMR signal was measured at lower $H$ (350, 500, 650 Oe), which resulted in $360\text{ deg}$-periodicity modulation. This anomalous periodicity could be due to the smooth periodic deviation of the $\vtr{F}$ and $\vtr{G}$ about its mean position during $\Hext$-rotation. In order to analyse the observed SMR, we first estimated the equilibrium spin-configuration by minimizing a simple Hamiltonian, included paramagnetic contribution of RE-sublattice and then, computed the SMR behaviour. The Hamiltonian was comprised of competing interactions acted upon \ch{Fe} spins; exchange, anisotropy, Zeeman and  Dzyaloshinskii-Moriya interactions. Our simulations showed close agreement with the experimental measured SMR.

In order to probe SRT ($\Gamma_{4} \xrightarrow{\Gamma_{1}} \Gamma_{2}$), SMR ($\alpha$ scans) under $H=2.4$ kOe was measured  at various temperatures down to 11 K. With lowering the temperature, the observed decrease in the width of the rotational hysteresis (down to 150 K) and a gradual reduction in the skewness of $\rhote$-modulation (in the range, 100 to 52 K) point towards the weakening of the anisotropy. Our SMR data (45 to 30 K) reflects the complexity pertaining to various phases ($\Gamma_{4}$, $\Gamma_{1}$ and $\Gamma_{2}$) and phase-transformations. Below 25 K, SMR signal showed sharp step-like feature around $c$-axis indicating realization of the $\Gamma_{2}$ phase where, $\vtr{F}\parallel \vtr{a}$ and $\vtr{G}\parallel \vtr{c}$. In summary, our studies highlight the relevance of SMR as a useful tool to track SRT as well as the anisotropy in 3D AFM and also underline the prospects of \ch{Ho_{0.5}Dy_{0.5}FeO_{3}} for use in AFM spintronic devices.

\section{ACKNOWLEDGEMENTS}

The authors acknowledge Department of Science and Technology, India for providing a project grant and are grateful to National Facility for Low Temperature and High Magnetic Field where electromagnetic and spin transport measurements were performed. Author, PG, thanks UGC India for research grant.

A.W. and P.G. contributed equally to this work.

\bibliography{HDFO}

\end{document}